# Phase-field insight into the nature of topological polarization structures in uniaxial ferroelectrics

Olga Y. Mazur[1,*], Pavel Marton[2,3], and Yuri Genenko[1]

[1]Institute of Materials Science, Technical University of Darmstadt, Otto-Berndt-Str, 3, 64287 Darmstadt, Germany
[2]Institute of Physics, Czech Academy of Sciences, Na Slovance 2, 182 00 Praha 8, Czech Republic
[3]Faculty of Mechatronics, Informatics and Interdisciplinary Studies, Technical University of Liberec, Studentská 2, 461 17 Liberec, Czech Republic
*mazur@mm.tu-darmstadt.de

**Abstract.** Uniaxial ferroelectrics have long served as benchmark systems for time-resolved studies of polarization evolution due to their simple 180° domain structure. However, recent experiments in these materials have revealed complex topological textures with enigmatic features, including the unexpected charge-free head-to-head and tail-to-tail domain walls which still wait for their consistent explanation. Here, we use a universal three-dimensional phase-field model to investigate the kinetics of domain structure development in triglycine sulfate, accounting for the monoclinic symmetry of its parent and ferroelectric phases. We show that isolated saddle-like polarization structures, interconnected saddle cascades, saddle lines and specific “monkey” saddles can form only when transverse degrees of freedom for spontaneous polarization are allowed in the polar plane of a uniaxial ferroelectric. The abundance and diversity of saddle-like structures at the early stages of domain ordering drives a far more effective compensation of bound charges at head-to-head and tail-to-tail domain walls than those predicted by single-component order parameter models. As domains coarsen and residual bound charges migrate toward the surfaces of the sample, the topological landscape simplifies, leaving behind only isolated saddle points. The present model successfully reproduces exotic structures observed experimentally in triglycine sulfate and lead germanate and reveals a domain topology that extends beyond the conventional Ising picture of uniaxial ferroelectrics. Although the mechanisms of bound-charge compensation are likely to be material-specific, similar topological structures may emerge in other uniaxial ferroelectrics with low spontaneous polarization and low Curie temperature. The general formulation of our phase-field model opens the way for future studies of such phenomena across a broader class of ferroelectric materials of arbitrary symmetry.

## 1. Introduction

Uniaxial ferroelectrics constitute a broad class of materials characterized by a unique (polar) axis along which oppositely oriented domains form. They are ideal materials for both theoretical and experimental studies for several reasons: typically, a single-phase transition from a high-temperature paraelectric phase to a low-temperature ferroelectric phase, a simple 180° domain structure, and the absence of ferroelasticity in some of them, e.g. triglycine sulfate (TGS), lead germanate (PGO), or guanidinium aluminum sulfate hexahydrate (GASH) [1]. Furthermore, beyond their use as model systems, uniaxial ferroelectrics find widespread application in devices: in highly sensitive infrared detectors based on TGS [2-4] and $LiTaO_3$ (LT) [5], high-speed electro-optic modulators [6,7], surface acoustic wave filters and laterally excited bulk-wave resonators based on $LiNbO_3$ (LN) and LT [8], multimodal neuromorphic devices based on $CuInP_2S_6$ [9], and others [10].

The functional performance of ferroelectric materials in electronic devices is largely determined by their domain structure, which governs important functional properties such as polarization switching, coercive field, dielectric and piezoelectric responses, and operational reliability. Over last four decades, TGS crystals have remained the definitive benchmark system for *in-situ* investigations into the formation and evolution of domain structures in uniaxial ferroelectrics. Its combination of readily available low-defect single crystals, a low Curie temperature and exceptionally high dielectric purity, arising from negligible electronic conductivity, slows down domain relaxation near the phase transition to timescales of hours or even days. This unique behavior enables direct experimental access to the fundamental mechanisms of domain nucleation, growth and ordering in the vicinity of the ferroelectric phase transition.

Early optical experiments employing liquid crystal coatings resolved the temporal evolution of TGS domain patterns across diverse thermal regimes, from the vicinity of Curie temperature ($0.03\ K < T_C - T < 3\ K$) [11] to the deep ferroelectric phase ($T_C - T = 24$ K) [12-14]. Further investigations were conducted using scanning force microscopy (SFM) [15-17], scanning electron microscopy (SEM), and atomic force microscopy (AFM) [19-22]. These studies mapped domain growth after quenching under an external electric field [14], thermal shock [22], irradiation, and doping [21], and consistently revealed conventional lamellar and lenticular 180° domains [23]. However, these techniques image only the top surface perpendicular to the spontaneous polarization $P_s$, obscuring the bulk dynamics. Since domains in TGS do not always traverse the entire sample thickness and domain walls can deviate from the polar axis [1,24], critical three-dimensional domain architecture remains hidden in the bulk. Such scenario, where most of the inverse domains end in the bulk of the crystal, implies a presence of appreciable bound electric charge distributed in the crystal, which is unfavorable from the electrostatic point of view.

First lateral-plane (parallel to the spontaneous polarization $P_s$) imaging using water etching and powdering revealed that these internal, non-through domains form elongated, cigar-shaped regions with elliptical cross-sections, tapered at both ends and aligned along the polar axis [25]. While such lateral profiles were also recorded afterward [26,27], direct imaging of surfaces parallel to the polar axis remains fundamentally restricted by a lack of $P_s$-normal bound surface charges, identical refractive index tensors preventing optical contrast in 180° domains, and inhibited differential chemical etching [24]. Early three-dimensional (3D) insights into the domain wall dynamics in TGS crystals were achieved by Wehmeier et al. via second-harmonic generation microscopy (SHGM) [28]. By tracking a 40-minute thermal cycle across the Curie temperature, this work captured the real-time evolution of domain nucleation and erasure. More recently, McCluskey et al. employed piezoresponse force microscopy (PFM) to map TGS domain patterns across both polar and lateral planes [29]. Subsequent 3D reconstruction of domain walls, accomplished via layer-by-layer AFM sectioning, revealed that bulk domain morphology deviates fundamentally from classic 180° structures, and consists of complex topological configurations featured by saddle points [29].

Analogous behavior has been documented in PGO crystals, prompting hypotheses that such saddle-point configurations resolve the anomalous charge neutrality observed at head-to-head (H-H) and tail-to-tail (T-T) domain walls [30,31]. Specifically, this absence of bound charges is postulated to be mediated by stochastically fluctuating transverse polarization components that remain energetically permissible despite the macroscopic uniaxiality of the crystal. Although phase-field simulations for these experiments seem to have captured charge compensation, the transverse components were incorporated so far merely as an induced background dielectric response rather than spontaneous polarization contributions [31]. Crucially, Fousek, Cross and Tagantsev demonstrated that a self-consistent Landau framework for uniaxial ferroelectrics strictly requires accounting for all three spontaneous polarization components, spanning the primary polar axis alongside the two in-plane degrees of freedom [1,32]. However, its phase-field implementation remains severely constrained by the low symmetry of the paraelectric parent phase of uniaxial ferroelectrics. Despite recent progress for trigonal ferroelectrics LN and LT [33-35], extending multi-component phase-field models to TGS remains substantially more challenging because of its monoclinic symmetry in both paraelectric and ferroelectric phases. To date, the experimental findings for TGS [29] stand only in qualitative agreement with the analytical predictions of reduced domain wall charges derived from the stochastic theory [36], however, a comprehensive numerical validation still remains absent.

Here we introduce a general phase-field framework that incorporates relevant energetic contributions without imposing symmetry-specific restrictions, making it applicable to arbitrary ferroelectrics. Utilizing this framework, we construct a thermodynamic Landau-Ginzburg-Devonshire (LGD) model for TGS, as conceptually proposed by Fousek et. al. [1,32], by explicitly integrating gradient, elastic, and electrostrictive energy contributions expressed in terms of a three-component polarization vector order parameter. The model enables for the first time realistic three-dimensional phase-field simulations of domain structures in TGS. We identify saddle points as predicted experimentally [29-31], quantify the local bound-charge in both polar and lateral planes, and reveal the role of the transverse components in charge compensation at H-H and T-T domain walls and in the formation of topological domain structures. The simulated domain patterns are in a good agreement with available experiments [14,21,29-31] and with the predictions of the stochastic theory [36-38].

## 2. Results

Triglycine sulfate $(NH_2CH_2COOH)_3H_2SO_4$ is a uniaxial monoclinic ferroelectric that undergoes second order phase transition of the order-disorder type at Curie temperature $T_C$ = 322 K with changing space group from $P2_{1/m}$ to $P2_1$. Upon cooling below $T_C$ the symmetry breaks down due to the loss of the mirror plane and creation of a long-range order by glycine molecule G1 on one side, producing the spontaneous polarization ($P_s$ = 0.022-0.028 C/m$^2$ at room temperature) along the *Y* direction or *b*-axis [1,39]. The domain structure forms through the slow growth of domains along the polar *Y* direction, resulting in needle-like shapes parallel to it, i.e. in *XY*-plan, also called lateral or nonpolar plane, and lenticular and elliptical shapes in the *XZ*-plane usually called polar plane [29,31].

The evolution of the TGS domain structure was investigated using the phase-field simulations in the framework of the LGD model with the dissipative Landau-Khalatnikov dynamics as implemented in simulation package FERRODO2 [40,41] and modified to accommodate arbitrary paraelectric parent phase symmetries. Our modeling is based on the LGD thermodynamic functional for TGS crystal that incorporates the polar component $P_y$ alongside the stochastically fluctuating in-plane components $P_x$ and $P_z$ (see the Phase-Field Model section for details). The simulation data were processed in MATLAB package (version R2026a). Figure 1 illustrates the simulated evolution of domain growth in TGS sample placed between two dielectric layers, replicating standard experimental boundary conditions [14,15,20,29]. Dielectric layers are placed on polar *XZ*-plane perpendicular to the direction of spontaneous polarization $P_y$ to account for electrically uncompensated surface. Hereinafter in the paper we will denote *XZ*-planes as bottom and upper related to the growing *Y* coordinate. The domain structure nucleates from an initial disordered state, with subsequent growth accompanied by extensive structural rearrangements.

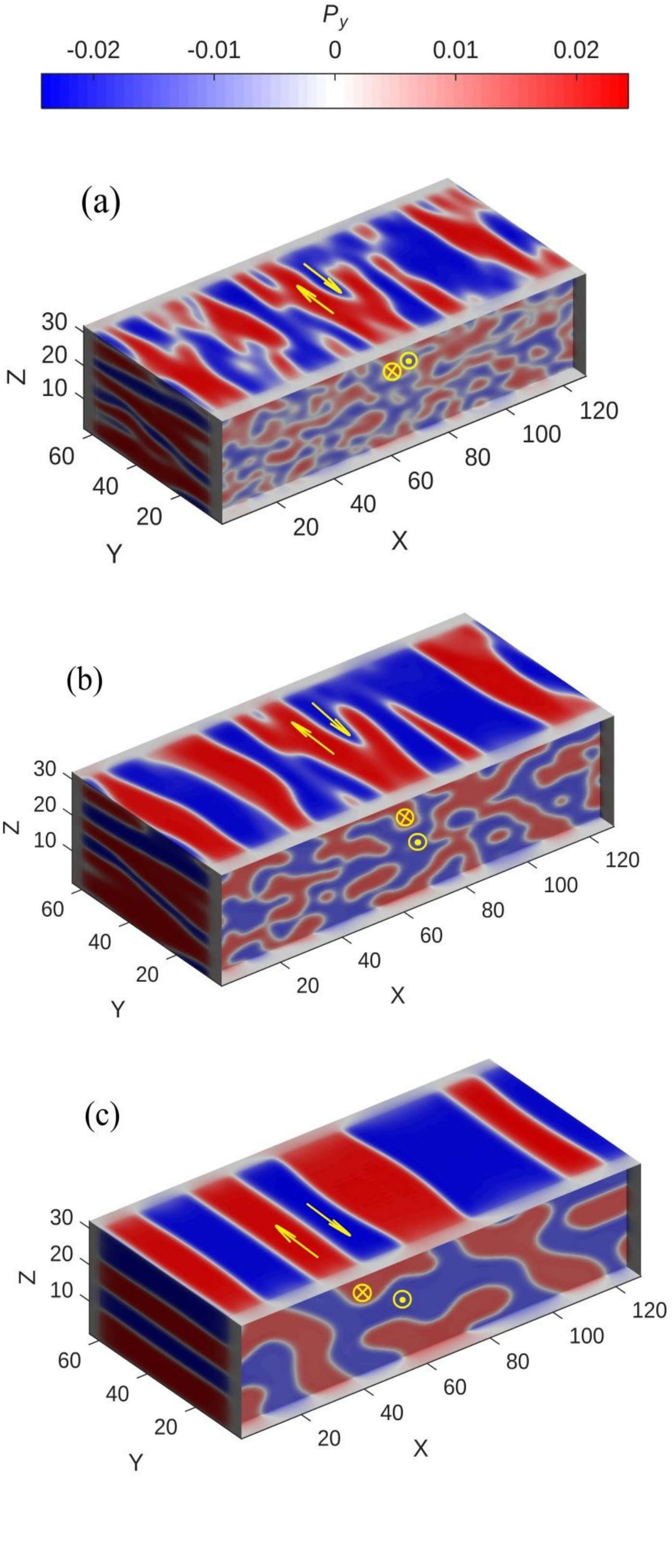

Fig. 1. Temporal evolution of the domain structure in TGS. Images (a), (b) and (c) correspond to the time moments of 4%, 9% and 100% of adjusted calculation time. The structural dimensions are mapped with a spatial resolution of 2 nm per point. Red and blue regions correspond to domains with opposite polarization orientation, whereas the narrow transition regions indicate domain walls. Spontaneous polarization is directed along *Y*-axis and showed by yellow arrows and circles. Value of spontaneous polarization is shown within colormap in C/m$^2$. Grey regions in *XZ*-plane correspond to the dielectric layers placed on the top and bottom of the sample.

The initial stage of the evolution is characterized by a large number of small domains, a high density of strongly curved domain walls, and the absence of a well-defined characteristic length scale (Fig. 1a). The *XZ* cross-section reveals pronounced structural heterogeneity throughout the bulk, whereas the surfaces of *XY*- and *YZ*-planes already exhibit the emergence of elongated domains preferentially aligned with the polar axis *Y*. Subsequent relaxation of the system lowers the free energy primarily through a reduction of the total domain-wall area. At the intermediate stage (Fig. 1b) this process results in a pronounced domain coarsening: small domains progressively disappear, while the transverse width of the stripe-like regions increases by several times. At the later long-lasting stage (Fig. 1c), coarsening continues and is accompanied by a gradual reduction in domain-wall curvature, although the bulk cross-section retains a labyrinthine morphology with numerous enclosed domains.

The surface structure becomes dominated by a small number of broad stripes, consistent with an increasing effective redistribution of polarization bound charge towards the free surfaces, aiming to organize the transverse configuration into parallel stripes. These observations indicate that domain coarsening proceeds slower in the bulk than at the surface. In reality, where defects should be accounted for, narrow domain walls with sharp polarization transitions severely restrict domain wall mobility. Consequently, a stripe-like pattern emerges on the *XY*- and *YZ*-planes, whereas the *XZ*-plane and bulk retain their labyrinthine architecture.

Three-dimensional reconstruction of domain walls across evolutionary stages provides direct insight into the internal structure, uncovering the formation of topological configurations driven by domain growth (Figs. 2-5). At the initial stage it reveals a highly corrugated and interconnected domain-wall network extending throughout the sample volume with domains hidden inside (Fig. 3). Rather than forming smooth, quasi-planar interfaces, the walls exhibit pronounced curvature, bifurcations and local connectivity, giving rise to a complex morphology. The reconstructed architecture is strongly heterogeneous in the sample volume, with elongated tortuous structures dominating along the *Y* direction, and more extended ridge-like forms emerging at the lower and upper boundaries (*XZ*-planes). Despite strong local distortions, extended wall segments retain preferential orientations, suggesting interplay between crystallographic anisotropy and the complex three-dimensional domain topology (Fig. 3a).

The domain architecture is primarily governed by three-dimensional saddle points and the associated saddle structures. Unlike a conventional two-dimensional saddle, where the polarization field converges along one direction and diverges along another, a three-dimensional saddle exhibits convergence along one axis and divergence along the two orthogonal directions, or vice versa [42]. In a uniaxial TGS crystal with the polar axis $P_y$, H-H and T-T domain walls are associated with two 3D saddle-point configurations (Fig. 2). At the H-H domain wall, the polarization field converges along *Y* direction through $P_y$ and diverges in the orthogonal plane through $P_x$ and $P_z$, corresponding to an $S_-$ saddle point (Fig. 2a). Conversely, at the T-T domain wall the polarization directions are reversed yielding an $S_+$ saddle point (Fig. 2b) [42].

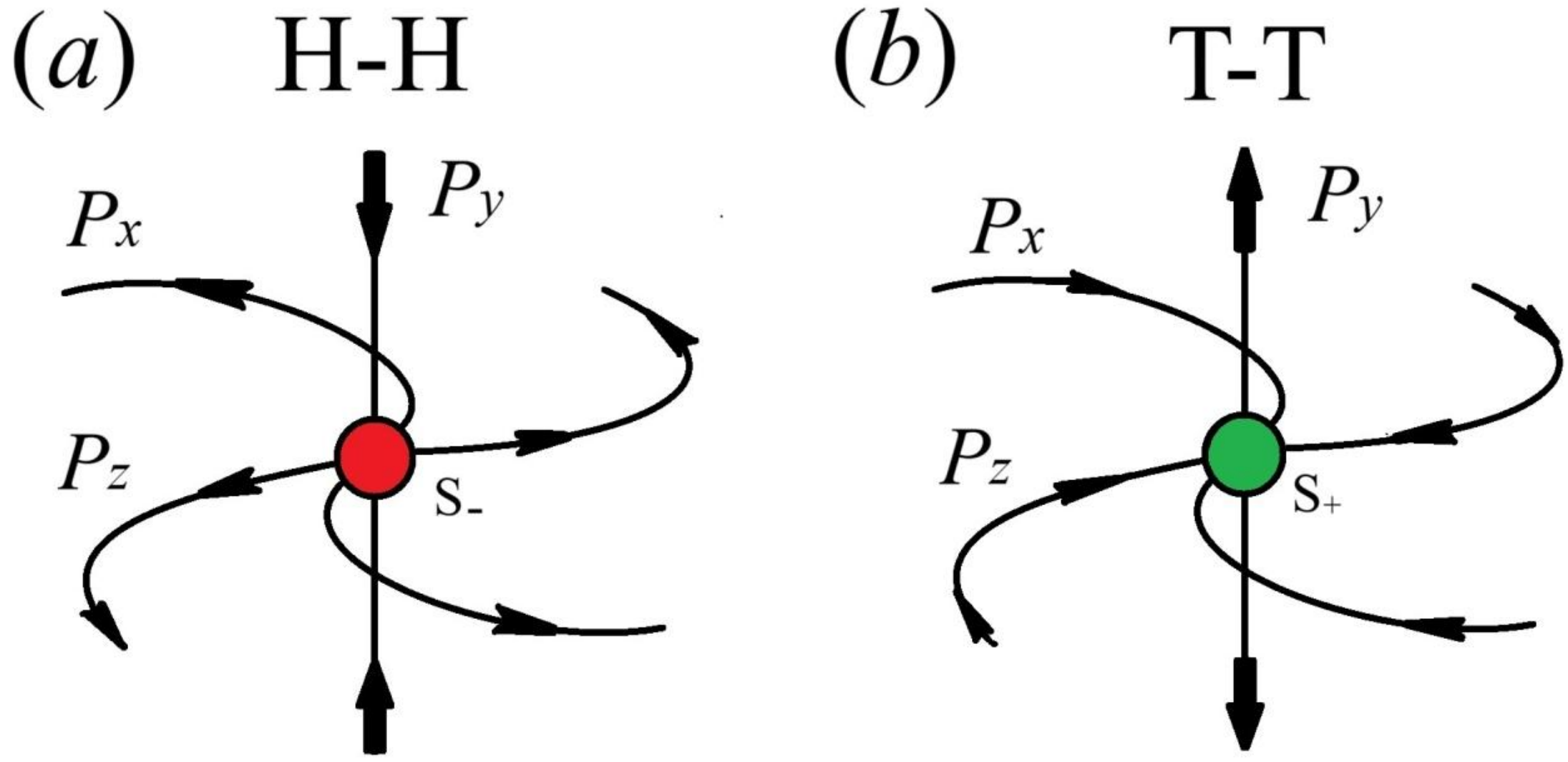

Fig. 2. Three-dimensional saddle-point configurations. (*a*) $S_-$ saddle point associated with a head-to-head (H-H) domain wall: the $P_y$ polarization component converges along the polar axis, while the transverse $P_x$ and $P_z$ components diverge. (*b*) $S_+$ saddle point associated with a tail-to-tail (T-T) domain wall, with $P_y$ divergence along the polar axis and transverse convergence of the $P_x$ and $P_z$ components [42]

Beyond the isolated saddle points characteristic of interlocking bifurcated domains, comprehensive mapping of the domain architecture reveals an extended network of interconnected $S_-$ and $S_+$ saddles, giving rise to complex branching structures (Fig. 3). Figures 3 (b,c) show an isolated $S_+$ saddle located at a T-T domain wall, where the transverse polarization components $P_x$ and $P_z$ converge towards the saddle point. The complementary configuration is observed at a H-H domain wall (Figs. 3d,e), where an $S_-$ saddle is formed. Consequently, displacement or annihilation of such a saddle point would require a collective reconfiguration of an extended segment of the domain wall. This non-local coupling between individual saddle points and the surrounding domain-wall morphology suggests that saddle structures may contribute to the emergence of collective order within the domain architecture.

The interconnected nature of the saddle structures becomes particularly pronounced within the bulk, where they develop into extended saddle surfaces comprising not only isolated and coupled saddle points but also saddle lines resembling the internal circumference of a torus, where each point is a saddle [42]. Figures 3 (f,g) show the domain-wall surface surrounding an internal domain fully enclosed within the crystal. In addition to a coupled structure formed by two saddle points $S_1$ and $S_2$, an extended saddle line $S_3$ emerges along the boundary of the enclosed domain. As the domain evolves, this line may progressively contract into a saddle point as the domain propagates along the polar axis. Alternatively, transverse displacement of the domain wall may transform the saddle line into an extended connection between neighboring saddle points, although this scenario is expected to be less favorable. Thus, saddle lines provide an intermediate structural motif linking localized saddle points to the extended three-dimensional topology of the domain-wall network.

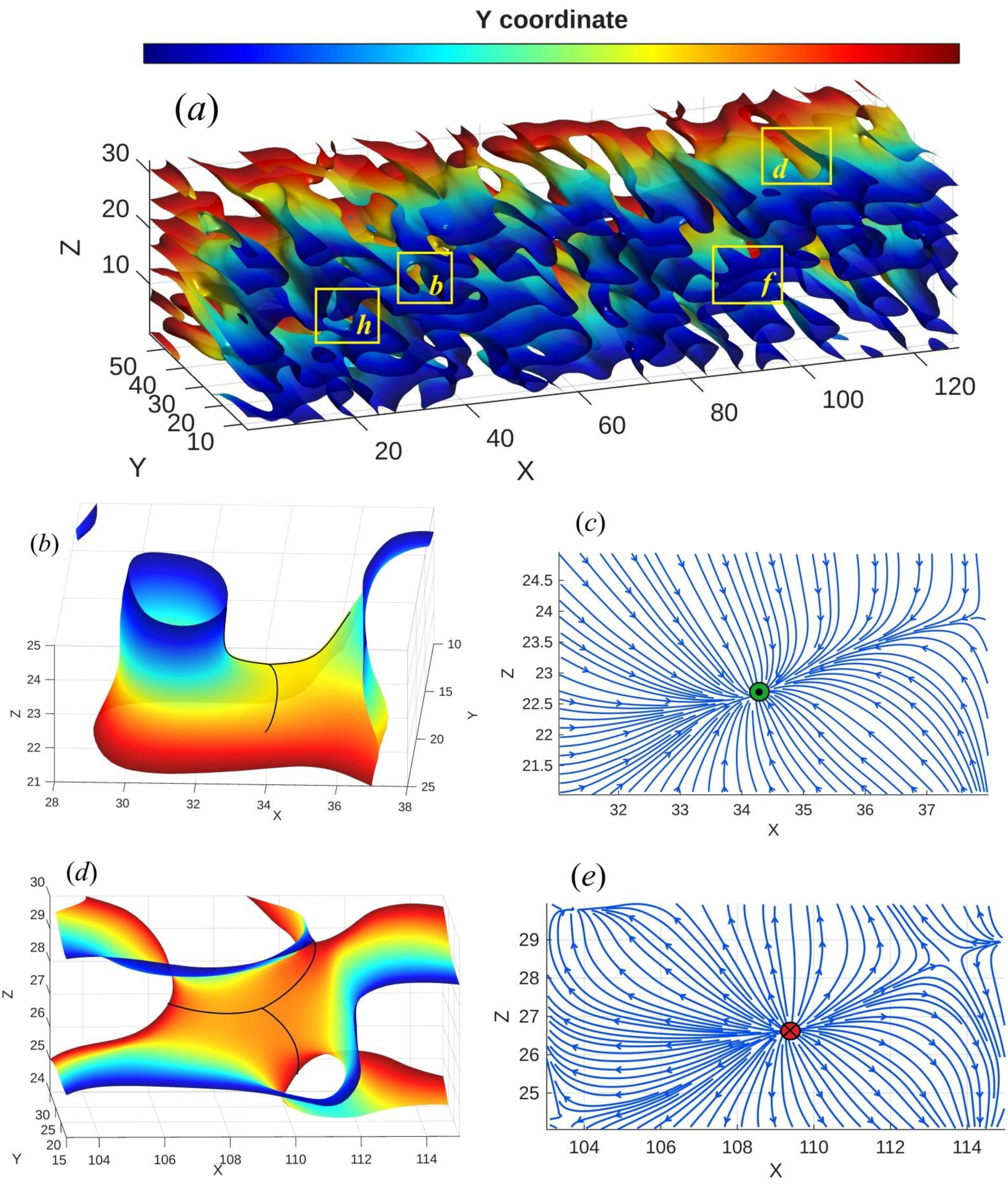

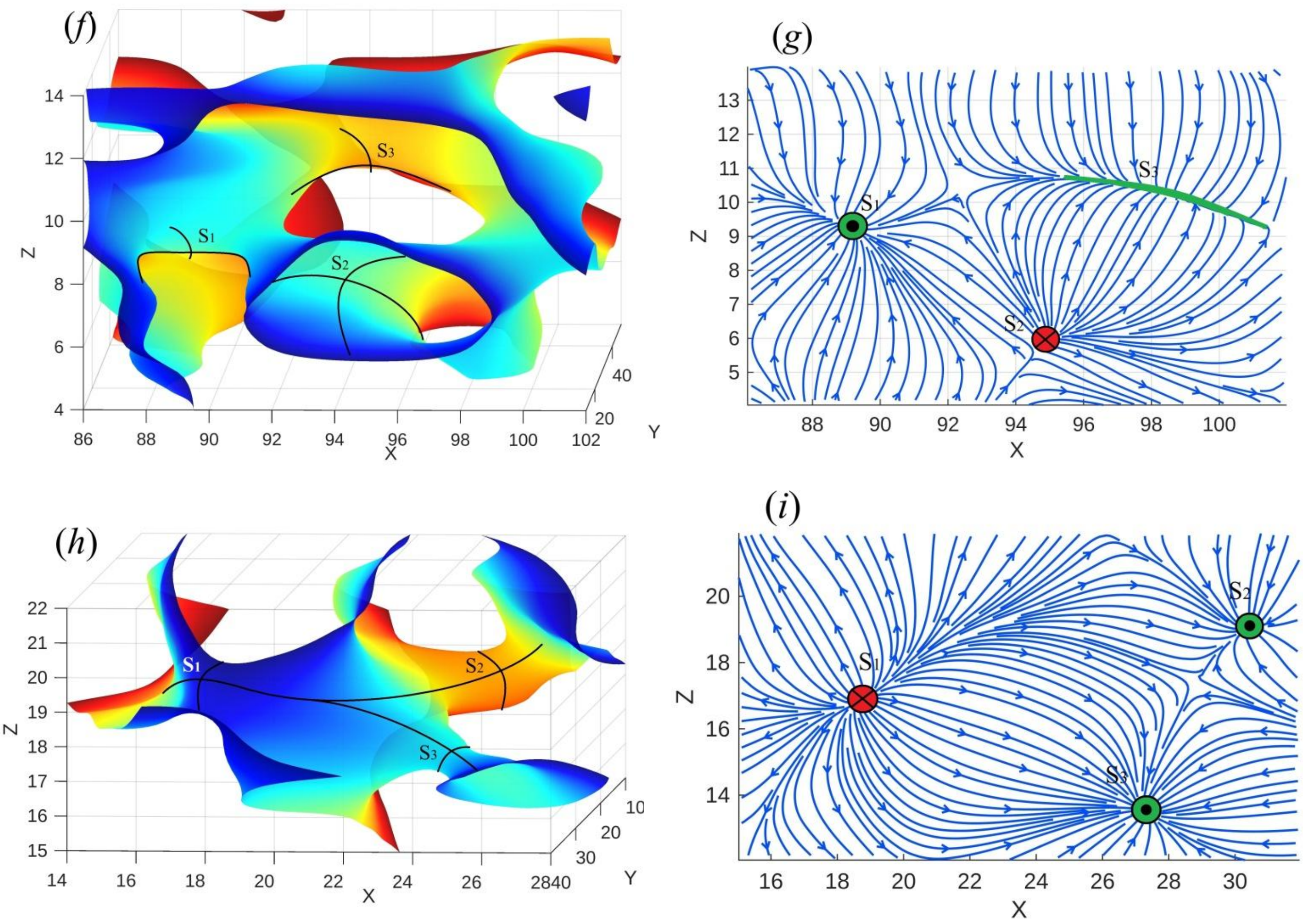

Fig. 3. Three-dimensional reconstruction of domain walls in TGS at the initial stage, corresponding to the domain structure shown in Fig. 1*a*. The color scale represents *Y* height, and the black lines show the curvature of the domain walls. The red circles with crosses correspond to saddle points $S_-$ formed on H-H domain walls. The green circles with points correspond to saddle points $S_+$ formed on T-T domain walls. (*a*) – Overview of the reconstructed domain-wall network. (*b*, *d*, *f*, *h*) – Magnified views of the regions highlighted in (*a*) with corresponding projection of 3D saddle configurations in *XZ*-plane (*c*, *e*, *g*, *i*) respectively. The visualizations in left column are rotated differently for better visibility. (*b*) – Isolated 3D saddle point. (*d*) – Three-branched saddle point structure. (*f*) – Saddle surface forming near the domain closed in the sample volume with the formation of two types of saddles $S_-$ ($S_1$) and $S_+$ ($S_2$) and saddle line $S_3$. (*f*) – Coupled "monkey" saddle structure formed by two "monkey" saddles created by connected pairs of saddle points $S_1$-$S_2$ and $S_1$-$S_3$. Streamlines represent the transverse polarization flow in *XZ*-plane: the divergence of lines from a single point indicates formation of $S_-$ 3D saddle point, while convergence indicates formation of $S_+$ 3D saddle point

An even more intricate topology emerges in the saddle surface shown in Fig. 3 (h,i). The $S_1$-$S_2$ and $S_1$-$S_3$ saddle-point pairs each define a surface known from the catastrophe theory as "monkey-saddle" [43,44]. Because the two monkey-saddle configurations share the common saddle point $S_1$ they collectively form a coupled monkey-saddle structure. Remarkably, each monkey saddle comprises saddle points of opposite type $S_-$ and $S_+$ and therefore bridges distinct H-H and T-T domain-wall configurations as is visualized in Fig. (3h,i). This coupling illustrates how locally distinct polarization singularities can become incorporated into a single three-dimensional domain-wall topology.

All these spatial configurations constitute a higher-order saddle network that mediates the connectivity between neighboring domain wall segments. Within these regions, the domain wall undergoes simultaneous multidirectional bending, geometrically coupling multiple segments. Consequently, the motion or reconfiguration of an individual saddle dictates a coordinated deformation of adjacent segments, driving collective rather than local domain wall dynamics. This coupling suppresses independent motion of isolated wall sections, giving rise to correlated regions that might govern the coercive field and macroscopic physical characteristics. The abundance of such complex, strongly interconnected saddle structures within the crystal bulk may impede the structural rearrangement of domain walls during relaxation. This topological constraint could account for the markedly different relaxation dynamics at the surface and in the bulk. While surface domains rapidly evolve towards a stripe-like configuration, the bulk domain network retains its complex three-dimensional morphology for considerably longer. This behavior is reflected in the persistence of the labyrinthine domain pattern in the polar plane even at late stages of relaxation (Fig. 1). We note that the mentioned topological features appear in a natural way from the initial disordered state after quenching and do not need application of special [31] boundary conditions for their creation.

The structural evolution of the system is governed by a relaxation process that drives network coarsening and pronounced orientation ordering (Fig. 4a). Highly curved initial elements attenuate as domain walls coalesce into extended continuous surfaces, minimizing excess area and local curvature (Fig. 4a), but with preserving some topological features (Fig 4b-e). This process yields a hierarchical morphology where lamellar-like order inside the volume coexists with a residual labyrinthine network close to the borders. Rather than complete evolution to a stripe-structure, relaxation redistributes topological complexity, suppressing micro-roughness while preserving macroscopic bridges, junctions, connections and saddle configurations. During this evolution, complex higher-order saddle-like structures, such as “monkey saddle” surfaces, gradually dissolve (Fig. 4b). The structural bridges interconnecting individual saddle points undergo progressive thinning until annihilation (Fig. 4b,c), rendering the system increasingly prone to the formation of isolated saddles (Fig. 4e,d). As the domains grow, the saddle-like structures connecting them progressively contract. Once the smaller domain is engulfed by the larger one, the associated saddle point is ultimately annihilated (Fig. 4d). The resulting state constitutes a partially ordered, three-dimensional domain-wall network characterized by two distinct scales: extended sheets interconnected by thin structural bridges (Fig. 4).

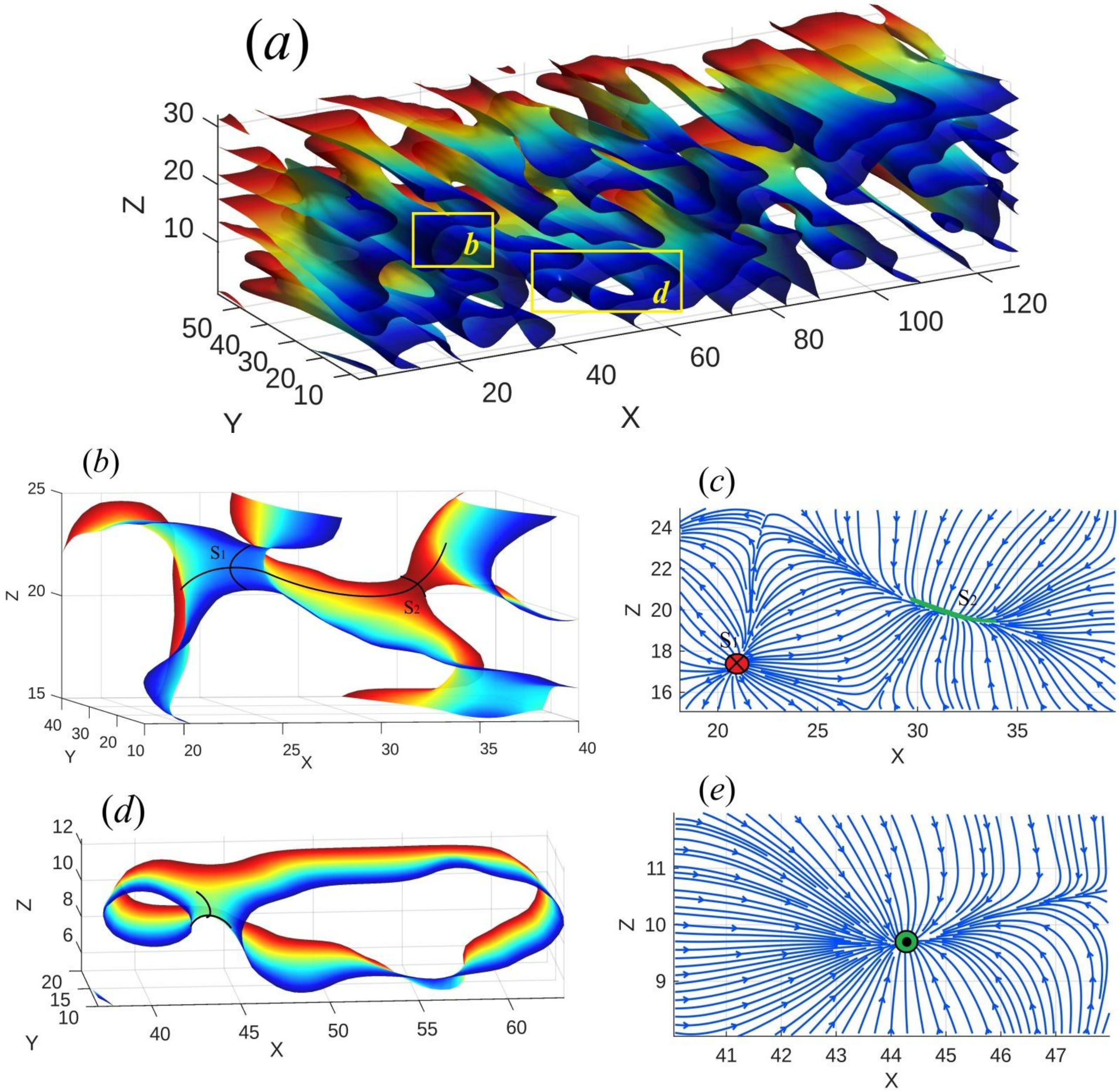

Fig. 4. Three-dimensional reconstruction of domain walls in TGS at the intermediate stage, corresponding to the domain configuration shown in Fig. 1*b*. (*a*) – Overview of the reconstructed domain-wall network. (*b*, *d*) – Magnified views of the regions highlighted in (*a*) with corresponding projection of 3D saddle configurations in *XZ*-plane (*c*, *e*) respectively. (*b*) – Lengthening of the ligament between saddle point $S_1$ (H-H domain wall) and saddle line $S_2$ (T-T domain wall). (*c*) – Isolated T-T saddle point at a domain wall separating two unequal-sized domains. All notations are the same as in Fig. 3

The final evolution stage represents the pronounced coarsening and topological simplification inside the sample (Fig. 5). The defining hallmark of this state is the near-complete collapse of the initial highly interconnected domain-wall network, which is replaced by a sparse configuration of large, spatially segregated, and smooth interfaces. This relaxation is driven not merely by the formation of stripe-like structure, but by a critical loss of network connectivity, marked by the annihilation of structural bridges, branching, and complex saddle-mediated junctions. This regime is characterized by three prominent features. First, the topological complexity drops drastically as the surfaces transition into predominantly simply-connected sheets or ribbons with large radii of curvature, effectively suppressing small-scale roughness. Second, the system retains anisotropy, because the remaining domain walls preferentially align along directional axes, as evidenced by the elongated sheets. Finally, the remaining topological features, such as residual neck-like configurations and sign-changing curvatures, become rare and spatially isolated (Fig. 5), contrasting sharply with the early-stage cascades (Figs. 3).

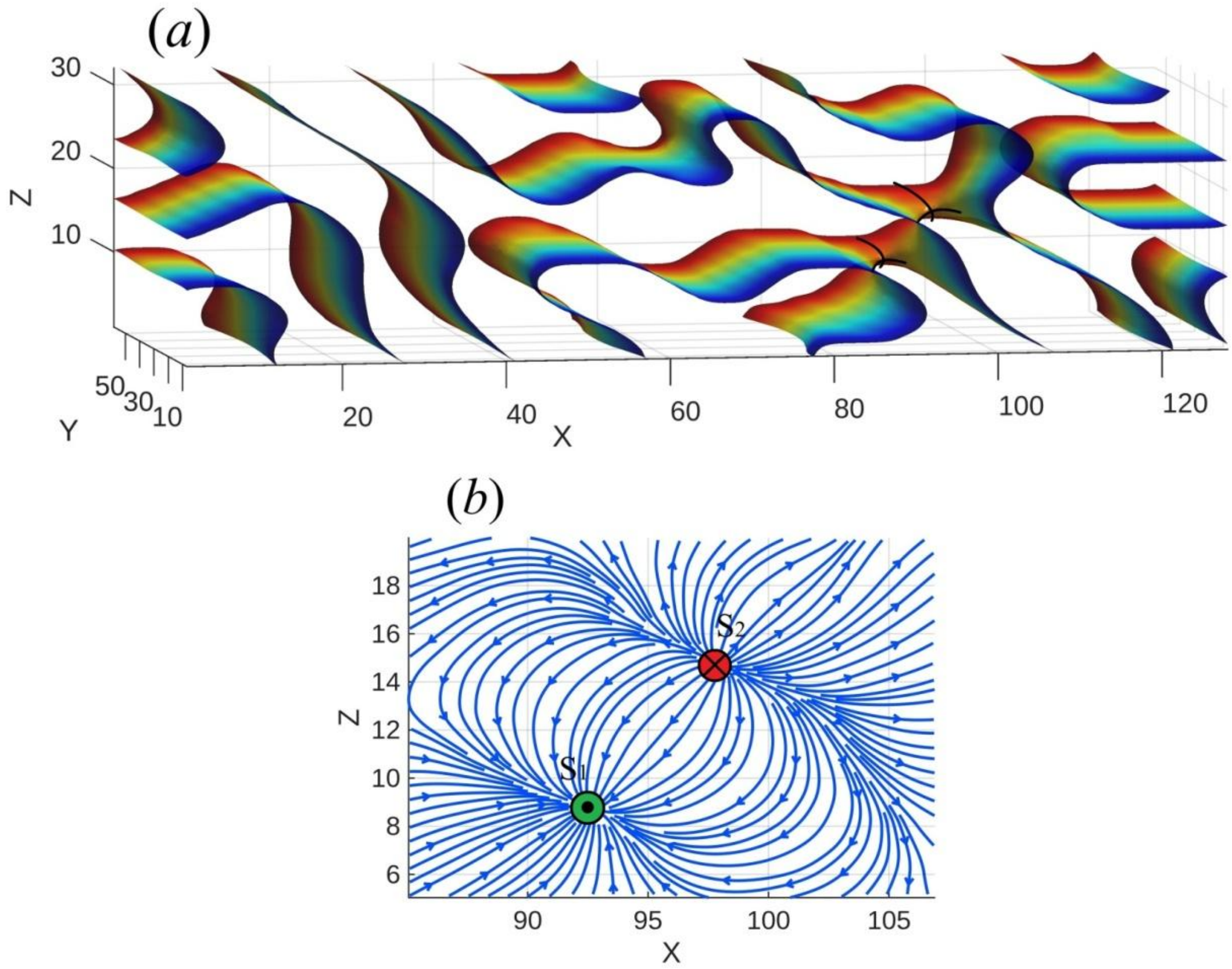

Fig. 5. Three-dimensional reconstruction of domain walls in TGS at the final stage, corresponding to the domain configuration shown in Fig. 1*c*. (*a*) – Overview of the reconstructed domain-wall network with preserved saddle points. (*b*) – Projection of 3D saddle configurations in *XZ*-plane indicating two type of saddle points $S_1$ (T-T domain wall) and $S_2$ (H-H domain wall)

Variety of topological defects during the early stages of relaxation, followed by their annihilation as the system approaches thermodynamic equilibrium, strongly supports the hypothesis that this intricate domain wall geometry accommodates charge compensation during the formation of H-H and T-T domain walls [29-31]. Despite the rigid macroscopic uniaxial anisotropy of TGS crystals, a local rotation of the polarization vector occurs directly at the interface. Consequently, opposing domains do not form stripe-like, charged frontal boundaries, but interpenetrate via highly branched, wedge-like, or interdigitated structures. The polarization vector continuously rotates around singular points, effectively suppressing polarization divergence and preventing bound charge accumulation. Driven by this mutual domain bifurcation, the emergent charge density at H-H and T-T domain walls remains at least one order of magnitude lower than that theoretically predicted for classical Ising-type walls.

To investigate the relationship between the domain structure and its electrical properties, spatial distributions of electric charge were shown. Figures (6-8) demonstrate the maps of charges contributed by polar component of polarization $\rho_y = -\partial_y P_y$ and transverse components $\rho_x + \rho_z = -\partial_x P_x - \partial_z P_z$ resulting from the decomposition of the total charge $\rho_{tot} = -\operatorname{div}(\boldsymbol{P})$ in *XZ*- (a, b, c) and *XY*- (d, e, f) planes across central lateral and polar cross-sections, respectively. In the initial, highly heterogeneous state (Fig. 6), large local charge densities from individual polarization components are substantially counterbalanced, as the polar component contribution $\rho_y$ is largely compensated by transverse components $\rho_x + \rho_z$. This indicates that highly curved, 3D domain walls accommodate polarization gradients while minimizing the electrostatic energy penalty. Residual uncompensated charge remains localized at certain domain wall segments, demonstrating that charge compensation is spatially heterogeneous and intimately coupled to the local three-dimensional wall geometry (Fig. 3). During subsequent relaxation (Fig. 7), the charge distribution coarsens and becomes more spatially ordered, tracking the progressive simplification of the domain wall morphology. This evolution is driven by a collective reorganization of the three-dimensional polarization field into an electrostatically favorable configuration, extending charge compensation across entire wall segments. At the final stage (Fig. 8), the remaining bound charge strongly localizes, leaving the bulk nearly charge-neutral. This process accompanies the topological simplification of the domain-wall network (Fig. 5), reflecting a coupled geometrical and electrostatic relaxation wherein the system expels interior charges toward the surfaces neighboring to electrodes.

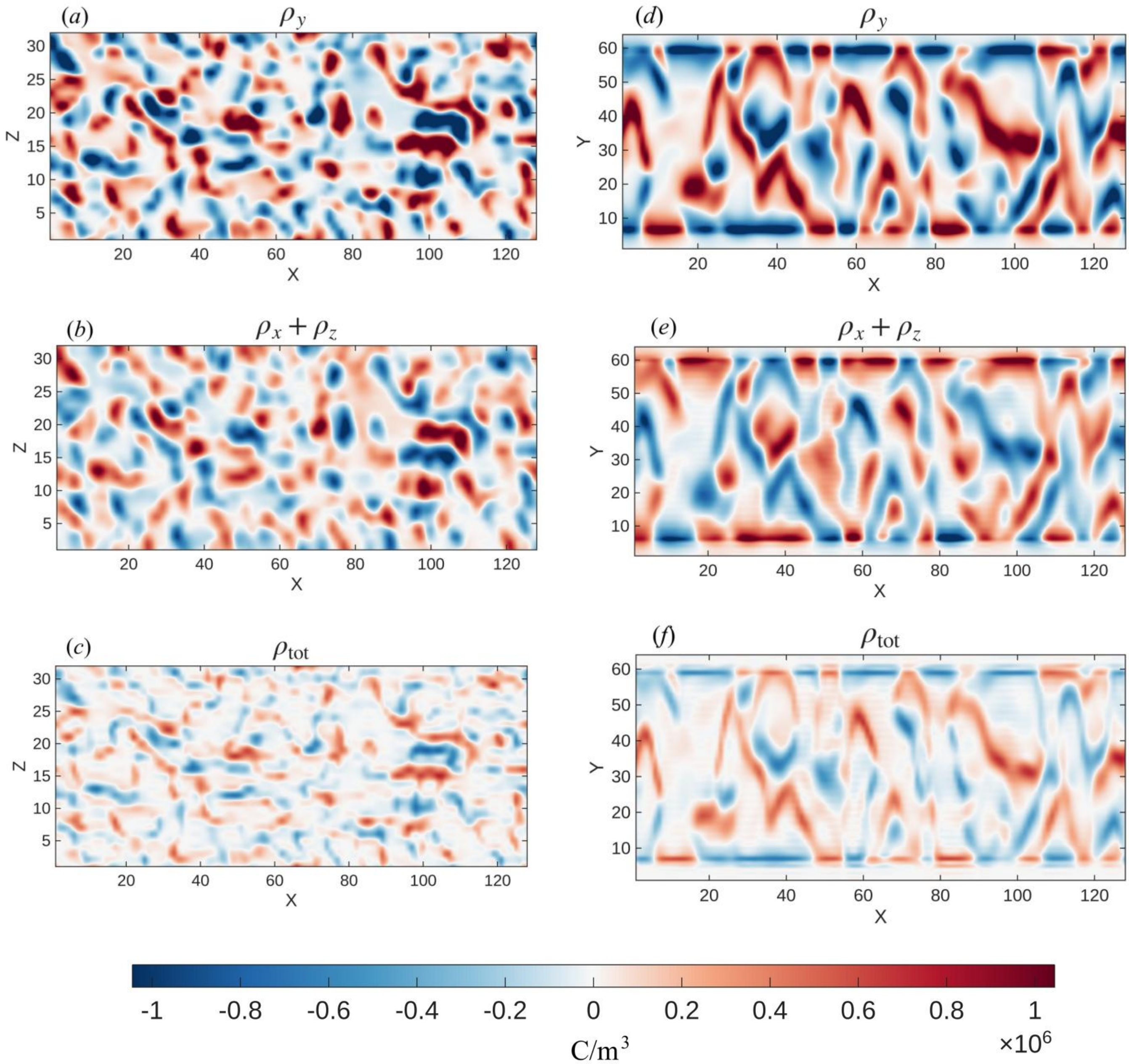

Fig. 6. Charge distribution maps, contributed by polarization components along polar axis $\rho_y = -\partial_y P_y$ (*a*, *d*), and in plane $\rho_x+\rho_z = -\partial_x P_x-\partial_z P_z$ (*b*, *e*), and total charge $\rho_{tot} = -\partial_y P_y-\partial_x P_x-\partial_z P_z$ (*c*, *f*) for the initial stage of relaxation, corresponding to the domain structure in Fig. 1*a*. Value of charge is shown within colormap in C/m$^3$. Figures *a*-*c* correspond to the polar *XZ*-plane in the central slice of the structure ($Y$ = 32), and figures *d*-*f* correspond to the lateral *XY* –plane in the central slice of the structure ($Z$ = 16)

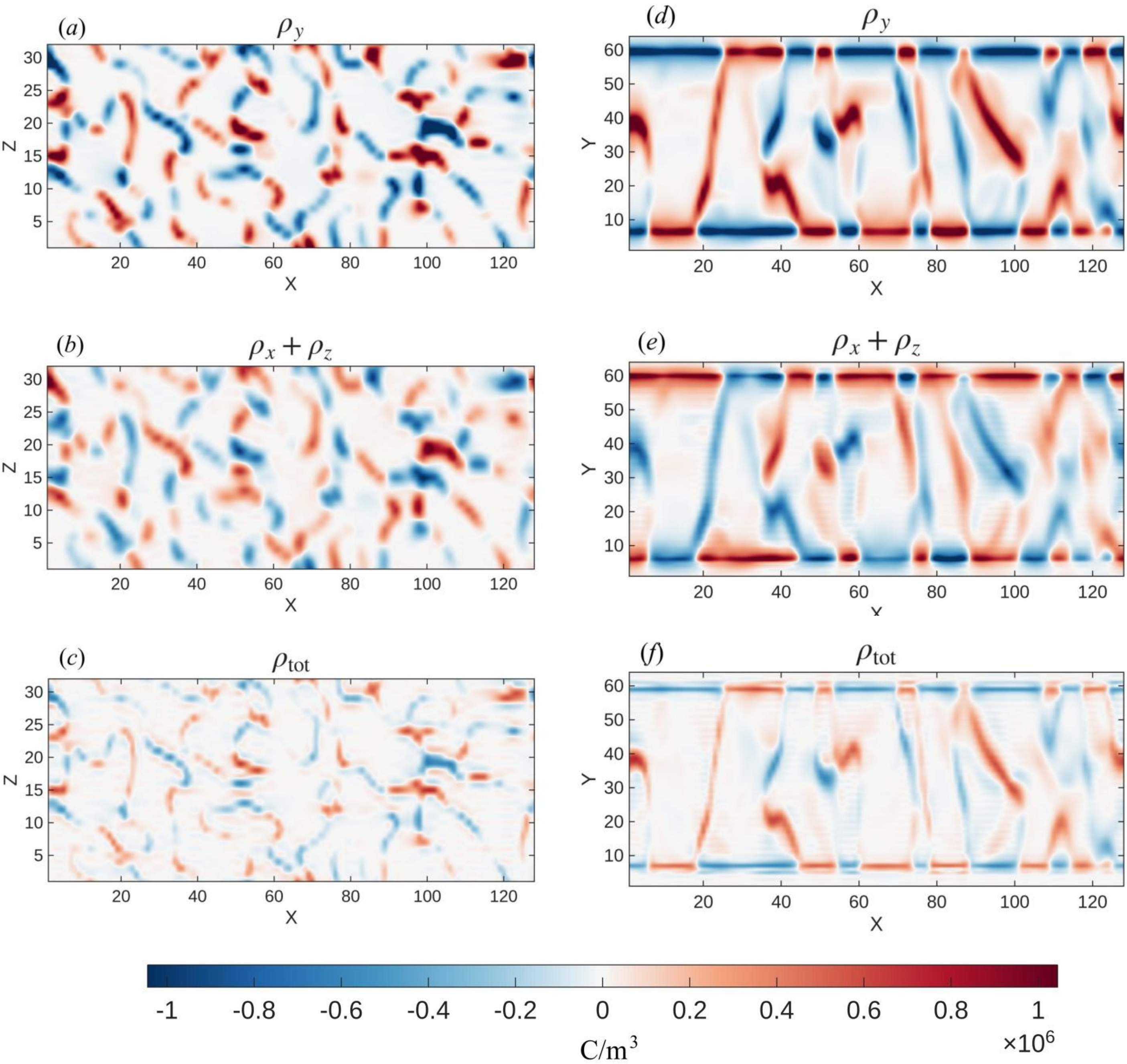

Fig. 7. Charge distribution maps, contributed by polarization components along polar axis $\rho_y = -\partial_y P_y$ (*a*, *d*) and in plane $\rho_x + \rho_z = -\partial_x P_x - \partial_z P_z$ (*b*, *e*), and total charge $\rho_{tot} = -\partial_y P_y - \partial_x P_x - \partial_z P_z$ (*c*, *f*) for the intermediate stage of relaxation, corresponding to the domain structure in Fig. 1*b*. Value of charge is shown within colormap in C/m$^3$. Figures *a*-*c* are correspond to the polar *XZ*-plane in the central slab of the structure ($Y$ = 32), and figures *d*-*f* correspond to the lateral *XY*-plane in the central slice of the structure ($Z$ = 16)

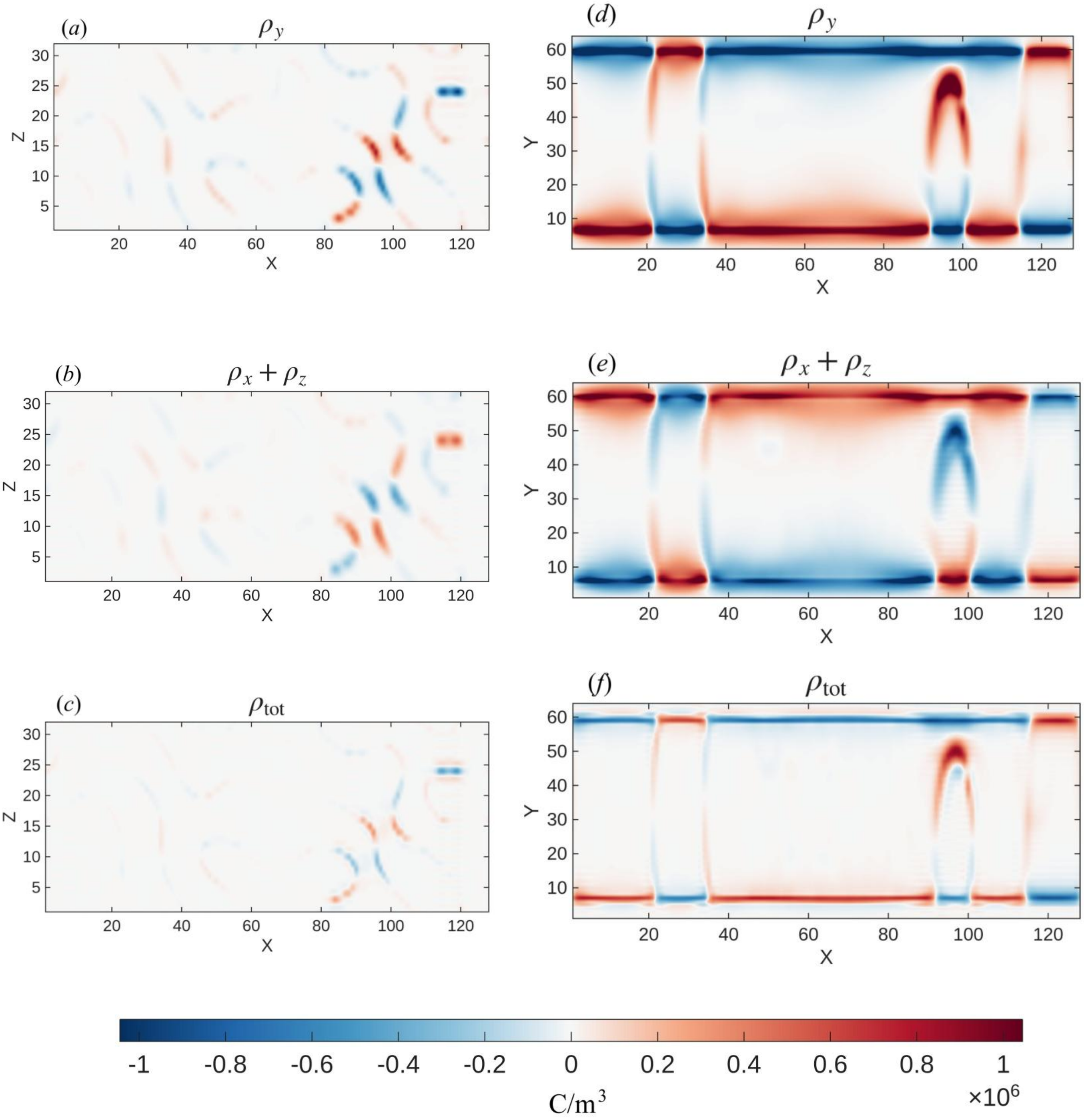

Fig. 8. Charge distribution maps, contributed by polarization components along polar axis $\rho_y = -\partial_y P_y$ (*a*, *d*) and in plane $\rho_x+\rho_z = -\partial_x P_x - \partial_z P_z$ (*b*, *e*), and total charge $\rho_{tot} = -\partial_y P_y - \partial_x P_x - \partial_z P_z$ (*c*, *f*) for the final stage of relaxation, corresponding to the domain structure in Fig. 1*c*. Value of charge is shown within colormap in $C/m^3$. Figures *a*-*c* are correspond to the polar *XZ*-plane in the central slab of the structure ($Y$ = 32), and figures *d*-*f* correspond to the lateral *XY*-plane in the central slice of the structure ($Z$ = 16)

## 3. Discussions

In general, any saddle point represents an inherently unstable state that defines a transient configuration during structural relaxation. In ideal conditions within a defect-free crystal, such topological features exist strictly during the nonequilibrium evolution of the system. In thermodynamic equilibrium, the driving force toward free energy minimization dictates that the domain network must transition completely into perfect 180° domain walls parallel to the polar direction. In real systems, however, even nominally pure TGS crystals, including those grown under microgravity conditions [45,46], possess a sufficient concentration of defects to stabilize these topological features throughout the bulk, preserving them as long-lived metastable configurations observed experimentally [29]. This multiscale roughness and connectivity highlight the intrinsically unusual character of the domain-wall configuration, which cannot be captured by a simple two-dimensional description within the Ising model of 180° domain walls.

The connection between saddle-point textures and bound-charge reduction is governed by the dimensionality of the polarization field. Within a single-component description, bound charge can be partially suppressed. Recently the analytical stochastic theory showed that spatial redistribution of the sole polar component broadens the transition between antiparallel domains, thereby reducing the local polarization gradient and the associated bound-charge density near nominally H-H and T-T configurations [36]. Allowing the polarization to explore all three components, as in the current research, introduces additional pathways for this redistribution. The transverse $P_x$ and $P_z$ components enable the polarization field to reorganize in three-dimensional order-parameter space rather than solely through variations of $P_y$, providing greater freedom to minimize the local polarization-field divergence and producing the spatially extended saddle-like textures observed here. Bound-charge reduction and saddle-point formation are therefore intimately connected, but they are not equivalent phenomena. The former can occur already in a single-component field, whereas the latter reflects the multidimensional topology accessible only when transverse polarization degrees of freedom are retained. This can be seen from the form of the Hessian matrix of second derivatives of polarization which allows identifying the type of topological structures (See Phase-Field model Section).

This distinction becomes particularly clear from the topology of the local free-energy landscape. In a single-component description, reversal between antiparallel domains is constrained to pass through ($P_y \sim 0$), which represents a maximum of the free energy along the switching coordinate. Once the transverse components are included, the corresponding configuration may instead acquire the character of a saddle point: unstable along the polar direction $Y$, yet locally stable with respect to perturbations along $X$ and $Z$. The appearance of such a multidimensional saddle does not, by itself, imply that transverse polarization components $P_x$ and $P_z$ lower the switching barrier. The reduced energetic cost may still originate predominantly from thermodynamic softening of the main polar component $P_y$, while the additional components merely reveal the transverse curvature of the free-energy surface. For modification of the energy barrier by $P_x$ and $P_z$ themselves, the free energy must contain sufficiently strong coupling between longitudinal and transverse polarization degrees of freedom, for example through higher-order mixed terms like $P_x^2 P_y^2$ . Such terms would allow transverse polarization to reshape the minimum-energy pathway, shifting its saddle away from the purely longitudinal trajectory and potentially lowering its energy. Establishing whether the observed saddle textures merely expose the multidimensional topology of the softened polar state, or actively provide a lower-energy switching pathway, therefore requires an explicit assessment of these higher-order couplings.

The emergence of strongly curved bulk domains and saddle-point polarization textures is not necessarily a universal feature of uniaxial ferroelectrics. Rather, the morphology adopted near H-H and T-T configurations is likely to reflect a material-specific competition between the thermodynamic cost of varying the ferroelectric order parameter and the electrostatic cost associated with its divergence. Importantly, the relative weight of these contributions cannot be inferred from the spontaneous polarization or bulk conductivity alone, as charge compensation at a domain wall may be governed by a complex redistribution of possibly present electronic and ionic charge carriers and defects.

Ferroelectrics LN and LT provide an instructive contrast to TGS and PGO crystals. In these materials, nominally charged domain walls can be stabilized retaining substantial polarization amplitude along the polar axis. The resulting bound charge is compensated by mobile carriers or charged defects. Gureev et al. showed within a Landau description that a periodic configuration of parallel H-H and T-T domain walls in an ideal defect-free proper ferroelectric is intrinsically unstable [47]. The symmetric charged domain configuration corresponds to a maximum of the energy, whereas in real crystals its position may be stabilized by defect-induced pinning [47]. The residual electrostatic cost can be further reduced through morphological reconstruction into stepped, faceted or zig-zag walls, for which local inclination reduces the normal component of the polarization discontinuity and hence the local bound-charge density [48-51]. Within this picture, the formation of a saddle-like polarization texture is not necessarily required for H-H or T-T configurations in LN and LT. Instead of eliminating the polar discontinuity through a three-dimensional reconstruction of the polarization field, these materials can sustain it as a screened and/or defect-stabilized charged wall [52]. Consistent with this interpretation, pronounced zig-zag and faceted charged-wall morphologies have been observed experimentally in lateral plane of LN [50], whereas saddle-point configurations analogous to those considered here have not, to our knowledge, been directly identified. Their existence cannot, however, be excluded, particularly deep in the bulk or on length scales inaccessible to conventional domain imaging techniques, and dedicated three-dimensional measurements would be required to establish whether such states are genuinely absent.

A second possible contribution may arise from differences in the underlying thermodynamic energy landscape. The large spontaneous polarization of LN and LT (~0.5–0.8 $C/m^2$) [53] is accompanied by a comparatively robust ferroelectric state. One may therefore hypothesize that local suppression or strong variation of the polar component of polarization carries a relatively high energetic penalty. By contrast, TGS and PGO have substantially smaller spontaneous polarizations (~0.023 and ~0.05 $C/m^2$ respectively) and may possess a softer local energy landscape with respect to polarization variation. Within the LGD description, such differences can be represented by the coefficients determining the depth and curvature of the ferroelectric free-energy minimum and the barrier separating symmetry-related polarization states. A shallower minimum or a lower barrier would make local suppression and reversal of the polarization less costly, potentially favoring continuous three-dimensional reorganization of the order parameter over the stabilization of a strongly charged H-H or T-T domain walls. We emphasize, however, that this interpretation remains a hypothesis:

quantitative comparison of the relevant Landau and gradient coefficients, screening energies and domain-wall energies across these materials would be required to establish whether thermodynamic "softness" is indeed responsible for the different domain configurations.

Ferroelectric PGO is particularly informative in this respect. Despite its semiconducting features and the associated availability of mobile charge carriers, it preferentially forms nominally uncharged H-H and T-T domain walls rather than the strongly charged ones observed in LN and LT. Instead, the preferred domain-wall configuration may reflect a competition between stabilizing a polar discontinuity through charge screening and reorganizing the polarization field to avoid such a discontinuity altogether. In PGO, the H-H or T-T configuration may instead evolve into a spatially extended saddle-like texture, suggesting that the absence of strongly charged walls reflects an energetically favorable reorganization of the polarization rather than an inability to screen bound charge. Another possible pathway may involve local antiferroelectric-like states at domain walls in PGO, as previously proposed [54].

Taken together, these considerations suggest that the contrast between LN/LT and TGS/PGO is unlikely to be controlled by a single material parameter. Charge-carrier availability, the energetics of polarization suppression and rotation, gradient energy, and the possible involvement of competing structural order parameters may all contribute. The experimentally observed zig-zag charged walls in LN and LT and saddle-like textures in the softer systems may therefore represent distinct solutions to the same electrostatic constraint. A quantitative comparison of these competing energy scales will be necessary to determine which mechanisms ultimately control the topology of domain configurations in each material.

On the basis of the physical considerations outlined above, namely, low spontaneous polarization, high structural susceptibility near room temperature and suppressed intrinsic carrier mobility, we anticipate that similar topological anomalies, bifurcated domain geometries and charge-free H-H and T-T domain walls may emerge in other soft uniaxial systems. Similar phenomena are expected to occur in ferroelectrics isomorphic to TGS, given their identical phase transition behavior and low Curie temperature, in particular, triglycine selenate TGSe ($T_C$ = 295 K, $P_s$ = 0.038 C/m$^2$) and triglycine fluoberyllate TGFB ($T_C$ = 346 K, $P_s$ = 0.037 C/m$^2$) [55]. Among prominent candidates are also Rochelle Salt ($T_C$ = 297 K, $P_s$ = 0.0025 C/m$^2$) and dicalcium strontium propionate DSP ($T_C$ = 281 K, $P_s$ = 0.003 C/m$^2$), and many others ferroelectrics with organic molecular structure [24]. Ferroelectrics GASH ($P_s$ = 0.0035 C/m$^2$) and tetramethylammonium trichloromercurate ($P_s$ = 0.01 C/m$^2$) which do not exhibit phase transitions at all, because Curie temperature is higher than their decomposition temperature, are of particular interest [24,56].

Among inorganic uniaxial ferroelectrics, novel topological features could be sought in crystals such as potassium nitrate $KNO_3$ ($T_C$ = 397 K, $P_s$ = 0.063 C/m$^2$) [57-59] and sodium nitrite $NaNO_2$ ($T_C$ = 433 K, $P_s$ = 0.064 C/m$^2$), both of which undergo a first order phase transition close to the second order [24]. Since characteristic first order features such as latent heat, delta-like heat capacity anomalies, and phase coexistence, are confined to a narrow temperature interval (usually $T-T_C$ < 1 K), these systems offer a viable platform for exploring domain structure kinetics. Potentially, analogous domain architectures could also manifest in the hexagonal multiferroic $InMnO_3$ ($T_C$ = 773 K, $P_s$ = 0.078 C/m$^2$) [60] and in the Van der Waals ferroelectric copper indium thiophosphate CIPS ($T_C$ = 292 K, $P_s$ = 0.04 C/m$^2$) [61]. By extension, materials characterized by high free-carrier concentrations, such as $KTiOPO_4$ (KTP), are anticipated to exhibit domain dynamics equivalent to those observed in LN and LT ferroelectrics.

Validating these topological phenomena across the suggested material spectrum should verify the universality of domain bifurcation in systems characterized by low spontaneous polarization and low transition temperatures. We note further that, of the whole variety of topological polarization structures which may be expected in ferroelectrics, such as skyrmions, hopfions and others [42], our simulations demonstrated only saddle points and combinations of them. The reason for this might be the uniaxiality of the studied TGS material. In contrast, simulations of multiaxial ferroelectric materials allow observation of topological structures like skyrmions in oxide superlatttices [62] and at the surface of thin films of lead zirconate titanate at high compressive misfit strains [63,64].

## 4. Conclusion

In summary, our phase-field simulations provide a thermodynamic and kinetic description of domain formation in the model uniaxial ferroelectric triglycine sulfate. By accounting for transverse polarization degrees of freedom, we demonstrate that the model system can dynamically mitigate polar discontinuities through the formation of complex, path-connected saddle-point geometries during early-stage growth. Under this model, the microstructural evolution is driven by the reduction of domain wall curvature, which gradually eliminates unstable saddle configurations. The low total bound charge observed at the domain wall ridges indicates that in the absence of extrinsic electronic screening, electrostatic neutrality can be maintained via intrinsic polarization rotation. This mechanism contrasts with more rigid lithium niobate and lithium tantalate, where highly charged zig-zag domain walls appear additionally stabilized by high free-carrier concentrations to screen polar discontinuities. Our results highlight the link between thermodynamic softness and the propensity to saddle-like textures, reflecting a balance between structural flexibility and electrostatic energy costs. These results reveal distinct, structurally driven pathways for accommodating uncharged H-H and T-T domain configurations without relying on free-carrier screening. Consequently, our findings provide a predictive basis for identifying non-Ising saddle structures in other structurally soft uniaxial ferroelectrics, such as TGS-isomorphs, Rochelle salt, potassium nitrate and others.

## 5. Phase-Field Model

The simulations were made in a phase-field model developed for general triclinic symmetry $C_1$ in parent phase based on classical Landau-Ginzburg-Devonshire theory, dissipative Landau-Khalatnikov dynamic equation, and Poisson equation, implemented in the code FERRODO2 and described in detail in [40,41,65]. In general form the model contains 94 terms in Landau energy $F_L$, 45 independent coefficients for gradient energy $F_G$, and 21 and 36 for elastic $F_C$ and electrosctriction $F_q$ energies, respectively, for the periodic boundary conditions. It is applicable to ferroelectrics with an arbitrary parent phase symmetry possessing an inversion operation. For TGS crystal the model was modified according to the form of energy contributions proposed in [1,32]:

$$F_L = \alpha_0(T - T_C)P_2^2 + \beta P_2^4 + \alpha_1 P_1^2 + \alpha_3 P_3^2 + \alpha_{13} P_1 P_3 \quad (1)$$

$$F_G = \frac{1}{2} G_{ijkl} P_{i,j} P_{k,l} = \frac{1}{2}\left(G_{2121} P_{2,1}^2 + G_{2222} P_{2,2}^2 + G_{2323} P_{2,3}^2 + G_{2123} P_{2,1} P_{2,3}\right) \quad (2)$$

$$F_C = \frac{1}{2} C_{ijkl} e_{ij} e_{kl} =$$
$$\frac{1}{2}(C_{1111} e_{11}^2 + C_{2222} e_2^{22} + C_{3333} e_{33}^2 + C_{2323} e_{23}^2 + C_{1313} e_{13}^2 + C_{1212} e_{12}^2)$$
$$+C_{1122} e_{11} e_{22} + C_{1133} e_{11} e_{33} + C_{2233} e_{22} e_{33} + C_{1113} e_{11} e_{13} + C_{1322} e_{13} e_2 + C_{1333} e_{13} e_{33} + C_{1223} e_{12} e_{23} \quad (3)$$

$$F_q = \frac{-1}{2} Q_{ijkl} e_{ij} P_k P_l =$$
$$-Q_{1122} e_{11} P_2^2 + Q_{2222} e_{22} P_2^2 + Q_{3322} e_{33} P_2^2 + Q_{1322} e_{13} P_2^2 + \quad (4)$$
$$2(Q_{2312} e_{23} P_1 P_2 + Q_{1212} e_{12} P_1 P_2 + Q_{1223} e_{12} P_2 P_3 + Q_{2323} e_{23} P_2 P_3)$$

Indices 1, 2, 3 correspond to the *x*, *y*, *z* in Cartesian coordinates; $G_{ijkl}$, $C_{ijkl}$, $Q_{ijkl}$ are gradient, elastic and electrostrictive tensors, respectively, where *i*, *j*, *k*, *l* take values from 1 to 3. Form of elastic tensor $C_{ijkl}$ is presented in [66]. Electrostriction tensor $Q_{ijkl}$ has the same form, but contains only terms with polar component $P_2$ of polarization vector. Values for the used coefficients are shown in Table 1.

The LGD functional (1) is sufficient to establish the existence of saddle points in the local free-energy landscape. The local curvature of the thermodynamic potential is determined by the Hessian matrix:

$$H = \frac{\partial^2 F_L}{\partial P_i \partial P_j} = \begin{pmatrix} 2\alpha_1 & 0 & \alpha_{13} \\ 0 & 2\alpha_0(T - T_C) + 12\beta P_2^2 & 0 \\ \alpha_{13} & 0 & 2\alpha_3 \end{pmatrix} \quad (5)$$

When the order parameter is restricted to the single polar component $P_2$, the relevant curvature at $P_2 = 0$ and $T < T_C$ satisfies $\partial_y^2 P_y^2 < 0$. Thus, $P_2 = 0$ corresponds to a local maximum along the polar switching coordinate. When all three polarization components are retained, however, the character of the same stationary point must be determined from the eigenvalues of the full Hessian. At $P_1 = P_2 = P_3 = 0$ and $T < T_C$, the curvature along the $P_2$ direction is negative, whereas the curvatures associated with the $P_1$ and $P_3$ directions can remain positive. The Hessian therefore has eigenvalues of opposite sign, identifying $P = 0$ as a saddle point of the three-dimensional local free-energy landscape rather than a simple maximum.

Importantly, in the form of the LGD functional (1), the main polar component $P_2$ is not energetically coupled to the transverse components $P_1$ and $P_3$. The inclusion of these additional degrees of freedom therefore changes the dimensional character of the stationary point but does not, by itself, modify the height of the polarization reversal barrier. Additional coupling terms, allowed by symmetry, such as $P_1^2 P_2^2$ or $P_1 P_2^3$, would couple the longitudinal and transverse polarization degrees of freedom and could consequently reshape both the free-energy landscape and the Hessian matrix (5). Such higher-order couplings can be incorporated into the present phase-field framework once a reliable parameterization of the corresponding LGD coefficients becomes available.

The computation domain was a rectangular box with dimensions 128 × 64 × 32 discrete points, where the dielectric layers of 5 points were placed on the top and on the bottom of the sample on the *XZ*-plane. Landau coefficients for dielectric layers were chosen as $\alpha_1 = 10^9$ J·m/C$^2$, $\alpha_2 = 2\cdot10^9$ J·m/C$^2$, $\alpha_3 = 10^9$ J·m/C$^2$. Since the thickness of the domain wall in TGS is evaluated in a range of 2-3.5 nm [1], the discretization step was chosen as $\Delta = 2$ nm. Estimating the domain wall thickness using formula (23) in [67] also yielded a value of 3.575 nm. Additional simulations with a larger discretization $\Delta = 3.5$ nm show similar results without complete pinning effects, but with slower mobility of domain walls and longer relaxation to the thermodynamically stable state. The simulations were conducted in mechanically free conditions, starting from initial state with $|P_0| = 0.001$ C/m$^2$ randomly distributed within the sample.

| $F_L$ | | $G_{ijkl}$, $\cdot10^{-9}$, J·m$^3$/C$^2$ | | $C_{ijkl}$, GPa | | $Q_{ijkl}$, m$^4$/C$^2$ | |
|---|---|---|---|---|---|---|---|
| $\alpha_0$ | $3.5\cdot10^7$ J·m/C$^2$ | $G_{2121}$ | 3.33 | $C_{1111}$ | 40.7 | $Q_{1122}$ | 1.3 |
| $\beta$ | $7.5\cdot10^{11}$ J·m$^5$/C$^4$ | $G_{2222}$ | 3.24 | $C_{2222}$ | 32.4 | $Q_{2222}$ | 2.2 |
| $\alpha_1^{eff}$ | $1.5\cdot10^9$ J·m/C$^2$ | $G_{2323}$ | 1.197 | $C_{3333}$ | 28.4 | $Q_{2233}$ | -3.9 |
| $\alpha_3^{eff}$ | $2.4\cdot10^9$ J·m/C$^2$ | $G_{2123}$ | -0.243 | $C_{2323}$ | 9.46 | $Q_{1322}$ | -0.3 |
| $\alpha_{13}^{eff}$ | $-0.127\cdot10^9$ J·m/C$^2$ | | | $C_{1313}$ | 11.4 | $Q_{2323}$ | -3.8 |
| | | | | $C_{1212}$ | 5.25 | $Q_{1223}$ | 0.5 |
| | | | | $C_{1122}$ | 16.0 | $Q_{2312}$ | 3.7 |
| | | | | $C_{1133}$ | 20.1 | $Q_{1212}$ | 1.9 |
| | | | | $C_{2233}$ | 16.8 | | |
| | | | | $C_{1113}$ | 1.7 | | |
| | | | | $C_{2213}$ | -2.1 | | |
| | | | | $C_{3313}$ | 2.7 | | |
| | | | | $C_{1223}$ | -1.3 | | |

Table 1. Values for TGS crystal known from experimental data and used for simulations. Landau coefficients $\alpha_0$ and $\beta$ are given in [68]. Values for gradient tensor $G_{ijkl}$ are given in [1,69], and for elastic $C_{ijkl}$ and electrostrictive $Q_{ijkl}$ tensors in [70,71], where alternative sets of measured values are also presented, demonstrating negligible variation. Coefficients $\alpha_1$, $\alpha_3$, $\alpha_{13}$ are not known in the literature, and therefore were calculated from the known values of dielectric permittivity $\varepsilon_{11} = 8.6$, $\varepsilon_{22} = 7$, $\varepsilon_{33} = 5.7$, $\varepsilon_{13} = 0.4$ [24,66]. A bit different values $\varepsilon_{11} = 10$, $\varepsilon_{22} = 10$, $\varepsilon_{33} = 8$ are also given in [72]. Such calculation gives the order of $\alpha_1$, $\alpha_3$, $\alpha_{13}$ parameters as $10^{10}$, which was reduced for one order to appropriately describe experimental results. These parameters were indicated as effective coefficients $\alpha_1^{eff}$, $\alpha_3^{eff}$, $\alpha_{13}^{eff}$ and can be used until the precise determination of the Landau coefficients values within density functional theory or other methods.

**Acknowledgements**

Enlightening discussions of phase-field simulation implementation with Igor Lukyanchuk are gratefully acknowledged. The authors are deeply grateful to Petr Ondrejkovič for insightful guidance on phase-field data analysis. The authors highly appreciate the discussions with Karsten

Albe concerning the application of this approach to ferroelectrics with structural defects. O.Y. Mazur gratefully acknowledges the financial support from the Alexander von Humboldt Foundation, within the framework of which this research was conducted. PM also acknowledge funding provided by the Ferroic Multifunctionalities project, supported by the Ministry of Education, Youth, and Sports of the Czech Republic. Project No. CZ.02.01.01/00/22_008/0004591, co-funded by the European Union. The research was supported by Czech Science Foundation Grant No. 24-11275S. Computational resources were provided by the e-INFRA CZ project (ID:90254), supported by the Ministry of Education, Youth and Sports of the Czech Republic.